\batchmode
\documentstyle[12pt]{article}
\newcommand{\nc}{\newcommand}
\nc{\renc}{\renewcommand}

\nc{\half}{{\textstyle{1\over2}}}
\nc{\etal}{\mbox{\it et al. }}
\nc{\ie}{{\it i.e.}}
\nc{\eg}{{\it e.g.}}

\renc{\thefootnote}{\arabic{footnote}}
\nc{\capt}[1]{{\bf Figure.} {\small\sl #1}}

\nc{\eqs}[2]{\mbox{Eqs.~(\ref{#1},\,\ref{#2})}}
\nc{\eq}[1]{\mbox{Eq.~(\ref{#1})}}

\nc{\figs}[2]{\mbox{Figs.~(\ref{#1},\,\ref{#2})}}
\nc{\fig}[1]{\mbox{Fig~.(\ref{#1})}}

\nc{\tag}[1]{\label{#1} \marginpar{{\footnotesize #1}}}
\nc{\mtag}[1]{\label{#1} \mbox{\marginpar{{\footnotesize #1}}}}
\renc{\baselinestretch}{1.5}
\newlength{\overeqskip}
\newlength{\undereqskip}
\nc{\be}[1]{\begin{equation} \mbox{$\label{#1}$}}
\nc{\bea}[1]{\begin{eqnarray} \mbox{$\label{#1}$}}
\nc{\Section}[2]{\section{#2}\label{#1}}
\nc{\Bibitem}[1]{\bibitem{#1}}
\nc{\Label}[1]{\label{#1}}

\nc{\eea}{\vspace{\undereqskip}\end{eqnarray}}
\nc{\ee}{\vspace{\undereqskip}\end{equation}}
\nc{\bdm}{\begin{displaymath}}
\nc{\edm}{\end{displaymath}}
\nc{\dpsty}{\displaystyle}
\nc{\bc}{\begin{center}}
\nc{\ec}{\end{center}}
\nc{\ba}{\begin{array}}
\nc{\ea}{\end{array}}
\nc{\bab}{\begin{abstract}}
\nc{\eab}{\end{abstract}}
\nc{\btab}{\begin{tabular}}
\nc{\etab}{\end{tabular}}
\nc{\bit}{\begin{itemize}}
\nc{\eit}{\end{itemize}}
\nc{\ben}{\begin{enumerate}}
\nc{\een}{\end{enumerate}}
\nc{\bfig}{\begin{figure}}
\nc{\efig}{\end{figure}}
\nc{\arreq}{&\!=\!&}
\nc{\arrmi}{&\!-\!&}
\nc{\arrpl}{&\!+\!&}
\nc{\arrap}{&\!\!\!\approx\!\!\!&}
\nc{\non}{\nonumber\\*}
\nc{\align}{\!\!\!\!\!\!\!\!&&}

\def\lsim{\; \raise0.3ex\hbox{$<$\kern-0.75em
      \raise-1.1ex\hbox{$\sim$}}\; }
\def\gsim{\; \raise0.3ex\hbox{$>$\kern-0.75em
      \raise-1.1ex\hbox{$\sim$}}\; }
\nc{\DOT}{\hspace{-0.08in}{\bf .}\hspace{0.1in}}
\nc{\Laada}{\hbox {$\sqcap$ \kern -1em $\sqcup$}}
\nc\loota{{\scriptstyle\sqcap\kern-0.55em\hbox{$\scriptstyle\sqcup$}}}
\nc\Loota{{\sqcap\kern-0.65em\hbox{$\sqcup$}}}
\nc\laada{\Loota}
\nc{\qed}{\hskip 3em \hbox{\BOX} \vskip 2ex}

\nc{\real}{{\rm I \! R}}
\nc{\Z}{{\sf Z \!\!\! Z}}
\nc{\complex}{{\rm C\!\!\! {\sf I}\,\,}}
\def\bigid{\leavevmode\hbox{\small1\kern-3.8pt\normalsize1}}
\def\id{\leavevmode\hbox{\small1\kern-3.3pt\normalsize1}}
\nc{\slask}{\!\!\!/}
\nc{\bis}{{\prime\prime}}
\nc{\pa}{\partial}
\nc{\na}{\nabla}
\nc{\ra}{\rangle}
\nc{\la}{\langle}
\nc{\goto}{\rightarrow}
\nc{\swap}{\leftrightarrow}

\nc{\EE}[1]{ \mbox{$\cdot10^{#1}$} }
\nc{\abs}[1]{\left|#1\right|}
\nc{\at}[2]{\left.#1\right|_{#2}}
\nc{\norm}[1]{\|#1\|}
\nc{\abscut}[2]{\Abs{#1}_{\scriptscriptstyle#2}}
\nc{\vek}[1]{{\rm\bf #1}}
\nc{\integral}[2]{\int\limits_{#1}^{#2}}
\nc{\inv}[1]{\frac{1}{#1}}
\nc{\dd}[2]{{{\partial #1}\over{\partial #2}}}
\nc{\ddd}[2]{{{{\partial}^2 #1}\over{\partial {#2}^2}}}
\nc{\dddd}[3]{{{{\partial}^2 #1}\over
	{\partial #2 \partial #3}}}
\nc{\dder}[2]{{{d #1}\over{d #2}}}
\nc{\ddder}[2]{{{d^2 #1}\over{d {#2}^2}}}
\nc{\dddder}[3]{{d^2 #1}\over
	{d #2 d #3}}
\nc{\dx}[1]{d\,^{#1}x}
\nc{\dy}[1]{d\,^{#1}y}
\nc{\dz}[1]{d\,^{#1}z}
\nc{\dl}[1]{\frac{d\,^{#1}l}{(2\pi)^{#1}}}
\nc{\dk}[1]{\frac{d\,^{#1}k}{(2\pi)^{#1}}}
\nc{\dq}[1]{\frac{d\,^{#1}q}{(2\pi)^{#1}}}

\nc{\cc}{\mbox{$c.c.$ }}
\nc{\hc}{\mbox{$h.c.$ }}
\nc{\cf}{cf.\ }
\nc{\erfc}{{\rm erfc}}
\nc{\Tr}{{\rm Tr\,}}
\nc{\tr}{{\rm tr\,}}
\nc{\pol}{{\rm pol}}
\nc{\sign}{{\rm sign}}
\nc{\bfT}{{\bf T }}

\def\GeV{{\rm\ GeV}}

\nc{\cA}{{\cal A}}
\nc{\cB}{{\cal B}}
\nc{\cD}{{\cal D}}
\nc{\cE}{{\cal E}}
\nc{\cG}{{\cal G}}
\nc{\cH}{{\cal H}}
\nc{\cL}{{\cal L}}
\nc{\cO}{{\cal O}}
\nc{\cT}{{\cal T}}
\nc{\cN}{{\cal N}}
\nc{\rvac}[1]{|{\cal O}#1\rangle}
\nc{\lvac}[1]{\langle{\cal O}#1|}
\nc{\rvacb}[1]{|{\cal O}_\beta #1\rangle}
\nc{\lvacb}[1]{\langle{\cal O}_\beta #1 |}
\nc{\bb}{\bar{\beta}}
\nc{\bt}{\tilde{\beta}}
\nc{\ctH}{\tilde{\cal H}}
\nc{\chH}{\hat{\cal H}}
\nc{\1}{\aa}
\nc{\2}{\"{a}}
\nc{\3}{\"{o}}
\nc{\4}{\AA}
\nc{\5}{\"{A}}
\nc{\6}{\"{O}}
\nc{\al}{\alpha}
\nc{\g}{\gamma}
\nc{\Del}{\Delta}
\nc{\e}{\epsilon}
\nc{\eps}{\epsilon}
\nc{\lam}{\lambda}
\nc{\om}{\omega}
\nc{\Om}{\Omega}
\nc{\ve}{\varepsilon}
\nc{\mn}{{\mu\nu}}
\nc{\k}{\kappa}
\nc{\vp}{\varphi}

\nc{\advp}[3]{{\it  Adv.\ in\ Phys.\ }{{\bf #1} {(#2)} {#3}}}
\nc{\annp}[3]{{\it  Ann.\ Phys.\ (N.Y.)\ }{{\bf #1} {(#2)} {#3}}}
\nc{\apl}[3]{{\it  Appl. Phys. Lett. }{{\bf #1} {(#2)} {#3}}}
\nc{\apj}[3]{{\it  Ap.\ J.\ }{{\bf #1} {(#2)} {#3}}}
\nc{\apjl}[3]{{\it  Ap.\ J.\ Lett.\ }{{\bf #1} {(#2)} {#3}}}
\nc{\app}[3]{{\it Astropart.\ Phys.\ }{{\bf #1} {(#2)} {#3}}}
\nc{\cmp}[3]{{\it  Comm.\ Math.\ Phys.\ }{{ \bf #1} {(#2)} {#3}}}
\nc{\cqg}[3]{{\it  Class.\ Quant.\ Grav.\ }{{\bf #1} {(#2)} {#3}}}
\nc{\epl}[3]{{\it  Europhys.\ Lett.\ }{{\bf #1} {(#2)} {#3}}}
\nc{\ijmp}[3]{{\it Int.\ J.\ Mod.\ Phys.\ }{{\bf #1} {(#2)} {#3}}}
\nc{\ijtp}[3]{{\it Int.\ J.\ Theor.\ Phys.\ }{{\bf #1} {(#2)} {#3}}}
\nc{\jmp}[3]{{\it  J.\ Math.\ Phys.\ }{{ \bf #1} {(#2)} {#3}}}
\nc{\jpa}[3]{{\it  J.\ Phys.\ A\ }{{\bf #1} {(#2)} {#3}}}
\nc{\jpc}[3]{{\it  J.\ Phys.\ C\ }{{\bf #1} {(#2)} {#3}}}
\nc{\jap}[3]{{\it J.\ Appl.\ Phys.\ }{{\bf #1} {(#2)} {#3}}}
\nc{\jpsj}[3]{{\it J.\ Phys.\ Soc.\ Japan\ }{{\bf #1} {(#2)} {#3}}}
\nc{\lmp}[3]{{\it Lett.\ Math.\ Phys.\ }{{\bf #1} {(#2)} {#3}}}
\nc{\mpl}[3]{{\it  Mod.\ Phys.\ Lett.\ }{{\bf #1} {(#2)} {#3}}}
\nc{\ncim}[3]{{\it  Nuov.\ Cim.\ }{{\bf #1} {(#2)} {#3}}}
\nc{\np}[3]{{\it  Nucl.\ Phys.\ }{{\bf #1} {(#2)} {#3}}}
\nc{\npps}[3]{{\it  Nucl.\ Phys.\ Proc.\ Suppl.\ }{{\bf #1} {(#2)} {#3}}}
\nc{\pr}[3]{{\it Phys.\ Rev.\ }{{\bf #1} {(#2)} {#3}}}
\nc{\pra}[3]{{\it  Phys.\ Rev.\ A\ }{{\bf #1} {(#2)} {#3}}}
\nc{\prb}[3]{{\it  Phys.\ Rev.\ B\ }{{{\bf #1} {(#2)} {#3}}}}
\nc{\prc}[3]{{\it  Phys.\ Rev.\ C\ }{{\bf #1} {(#2)} {#3}}}
\nc{\prd}[3]{{\it  Phys.\ Rev.\ D\ }{{\bf #1} {(#2)} {#3}}}
\nc{\prl}[3]{{\it Phys.\ Rev.\ Lett.\ }{{\bf #1} {(#2)} {#3}}}
\nc{\pl}[3]{{\it  Phys.\ Lett.\ }{{\bf #1} {(#2)} {#3}}}
\nc{\prep}[3]{{\it Phys.\ Rep.\ }{{\bf #1} {(#2)} {#3}}}
\nc{\prsl}[3]{{\it Proc.\ R.\ Soc.\ London\ }{{\bf #1} {(#2)} {#3}}}
\nc{\ptp}[3]{{\it  Prog.\ Theor.\ Phys.\ }{{\bf #1} {(#2)} {#3}}}
\nc{\ptps}[3]{{\it  Prog\ Theor.\ Phys.\ suppl.\ }{{\bf #1} {(#2)} {#3}}}
\nc{\physa}[3]{{\it  Physica\ A\ }{{\bf #1} {(#2)} {#3}}}
\nc{\physb}[3]{{\it  Physica\ B\ }{{\bf #1} {(#2)} {#3}}}
\nc{\phys}[3]{{\it Physica\ }{{\bf #1} {(#2)} {#3}}}
\nc{\rmp}[3]{{\it  Rev.\ Mod.\ Phys.\ }{{\bf #1} {(#2)} {#3}}}
\nc{\rpp}[3]{{\it Rep.\ Prog.\ Phys.\ }{{\bf #1} {(#2)} {#3}}}
\nc{\sjnp}[3]{{\it Sov.\ J.\ Nucl.\ Phys.\ }{{\bf #1} {(#2)} {#3}}}
\nc{\spjetp}[3]{{\it Sov.\ Phys.\ JETP\ }{{\bf #1} {(#2)} {#3}}}
\nc{\yf}[3]{{\it Yad.\ Fiz.\ }{{\bf #1} {(#2)} {#3}}}
\nc{\zetp}[3]{{\it Zh.\ Eksp.\ Teor.\ Fiz.\  }{{\bf #1}  {(#2)} {#3}}}
\nc{\zp}[3]{{\it Z.\ Phys.\ }{{\bf #1} {(#2)} {#3}}}
\nc{\ibid}[3]{{\sl ibid.\ }{{\bf #1} {#2} {#3}}}
\nc{\rf}[1]{(\ref{#1})}
\nc{\nn}{\nonumber \\*}
\nc{\bfB}{\bf{B}}
\nc{\bfv}{\bf{v}}
\nc{\bfx}{\bf{x}}
\nc{\bfy}{\bf{y}}
\nc{\vx}{\vec{x}}
\nc{\vy}{\vec{y}}
\nc{\oB}{\overline{B}}
\nc{\oI}{\overline{I}}
\nc{\oR}{\overline{R}}
\nc{\rar}{\rightarrow}
\nc{\ti}{\times}
\nc{\slsh}{\hskip-5pt/}
\nc{\sm}{Standard~Model~}
\nc{\MP}{M_{\rm Pl}}
\nc{\tp}{t_{\rm Pl}}
\nc{\ave}{\bar{E}}

\nc{\eff}{{\rm eff}}
\nc{\kk}{\vek{k}}
\nc{\pp}{{\rm p}}
\nc{\ga}{g_{a\gamma}}
\nc{\vv}{\\}
\nc{\eee}{{\bf E}}
\nc{\bbb}{{\bf B}}
\nc{\qcd}{T_{\rm QCD}}
\nc{\G}{\rm \ G}
\def\vec#1{{\bf #1}}

\def\lae{\;^{<}_{\sim} \;} \def\gae{\; ^{>}_{\sim} \;}

\def\ell{e^{c}LL}

\begin{document}
{\title{\vskip-2truecm{\hfill {{\small \\
	\hfill \\
	}}\vskip 1truecm}
{\LARGE  Inflaton Condensate Fragmentation in Hybrid Inflation Models}}
{\author{
{\sc  John McDonald$^{1}$}\\
{\sl\small Theoretical Physics Division,
University of Liverpool,
Liverpool L69 3BX, England}
}
\maketitle
\begin{abstract}
\noindent

       Inflation ends with the formation of a Bose condensate of inflatons.
We show that in hybrid inflation models this
 condensate is typically unstable with respect
 to spatial perturbations and can fragment to condensate lumps.
The case of D-term inflation is considered as an example and it is shown that
fragmentation occurs if $\lambda \gae 0.2g$, where $\lambda$ is the
superpotential coupling and $g$ is the $U(1)_{FI}$ gauge coupling.
Condensate fragmentation can result in an
 effective enhancement of inflaton annihilations over decays as the main mode of
reheating. In the case of D-term inflation models in which the Standard Model
 fields carry $U(1)_{FI}$ charges,
if condensate fragmentation occurs then reheating is dominated by
inflaton annihilations, typically resulting in 
the overproduction of thermal gravitinos.
Fragmentation may also have important consequences for SUSY
 flat direction dynamics and for preheating.

\end{abstract}
\vfil
\footnoterule
{\small $^1$mcdonald@sune.amtp.liv.ac.uk}

\thispagestyle{empty}
\newpage
\setcounter{page}{1}

\section{Introduction}

                A common feature of the cosmology of
particle physics models is the formation of
Bose condensates of scalar particles.
Examples include axion condensates \cite{ax}, condensates
 of squarks and sleptons along flat directions of the minimal
 supersymmetric (SUSY) standard model
 (MSSM) (Affleck-Dine condensates \cite{ad,drt,gherg}) and
inflaton condensates which form at the end
 of inflation and whose decay is responsible for
reheating the Universe \cite{eu}. It is usually
 assumed that the scalar particles in the
 condensate are
 non-interacting, corresponding to coherent
 oscillations in a purely $\phi^{2}$ potential.
However, in many cases this is not true. In the case of axions, deviation
 of the angular pseudo-Nambu Goldstone axion
potential from a pure $\phi^{2}$ potential implies an attractive force
 between the axions which results in the
 growth of spatial perturbations and the formation of
 axion miniclusters \cite{am}.
In the case of the Affleck-Dine condensate, deviation
 from the $\phi^{2}$ potential, either
 due to the flattening of the potential above
 the messenger field mass (gauge-mediated
 SUSY breaking \cite{k,ks}) or due to radiative
 corrections from gaugino loops
 (gravity-mediated SUSY breaking \cite{bbb1,bbb2})
 results the fragmentation of the condensate
 to form Q-balls \cite{ks,bbb1,bbb2,kawa}.
 Thus the conventional view
 of cosmological condensates as being
spatially homogeneous coherently oscillating scalar
 fields is not generally true.
In particular, when the potential is
 'flatter-than-$\phi^{2}$', meaning $Min(V(\phi)/\phi^{2})$
 is at $\phi \neq 0$ (with $V(0) = 0$),
the condensate is unstable with respect to
 spatial perturbations and fragments to non-topological solitons
which we will refer to as condensate lumps.

                 Here we consider the question of the stability
 of the inflaton condensate with
 respect to spatial perturbations and
 the consequences of its fragmentation.
  The most natural inflation models are hybrid
 inflation models \cite{hi}, which, unlike the
 case of single-field inflation models, allow
 inflation to occur without requiring couplings to be very small.
We will therefore focus on hybrid inflation models in the following.

          Although our results for inflaton condensate fragmentation can apply
to hybrid inflation models in general, we will focus
 on the case of SUSY hybrid inflation models \cite{fti,dti,lr}, which
 will allow us to illustrate the general phenomenon of inflaton condensate
 fragmentation whilst applying the results to a case of considerable
 interest. SUSY hybrid inflation models
 are either of the F-term \cite{fti} or D-term \cite{dti,kmr} type.
The most interesting are the D-term models, which
 can evade the so-called $\eta$-problem
i.e. the flatness of the inflaton potential in the
 presence of supergravity corrections \cite{lr}.
 We will therefore focus on D-term
 inflation, whilst presenting the results in a form
 that will allow them to be applied to other
 hybrid inflation models.

         Recently it has been shown that it is also
 possible for inflation to end via "tachyonic
 preheating" i.e. the rapid growth of spatial perturbations
of the inflaton field in the presence of a 
tachyonic potential
 \cite{tp,tp2,cpr}. 
The mode by which hybrid inflation ends (inflaton
 condensate fragmentation or tachyonic preheating)
 will be sensitive to the initial conditions 
 at the phase transition ending hybrid inflation, 
in particular the rate of roll of the homogeneous scalar field
 relative to the rate of growth of the spatial perturbations.
This requires a full analysis of the 
dynamics of the inflaton field,
including the effect of radiative corrections to the inflaton potential 
\cite{mb}. 
Since in this paper we wish to study the
 growth of spatial perturbations of a homogeneous hybrid 
inflation condensate in general, using
 D-term inflation as a particular example, we will
 assume throughout that a coherently oscillating
 scalar field condensate initially exists.
 
     The paper is organised as follows. In Section 2
 we review the D-term hybrid
 inflation model. In Section 3 we discuss
 condensate instability in hybrid inflation models.
 In Section 4 we consider the evolution of spatial
 perturbations of a coherently oscillating
 condensate. In Section 5 we apply the results
 to the case of D-term inflation. In Section 6
 we consider possible consequences of inflaton
 condensate fragmentation, in particular
the enhancement of annihilations as a mode of
 reheating. In Section 7 we comment on the relationship 
between tachyonic preheating and
 inflaton condensate fragmentation. In Section 8
 we present our conclusions.

 \section{D-term Hybrid Inflation}

         The superpotential of D-term inflation models is \cite{dti}
\be{e3} W = \lambda S \Phi_{+} \Phi_{-}  ~,\ee
resulting in a scalar potential
\be{e4} V = \lambda^{2} |S|^{2} (| \Phi_{+}|^{2} +
| \Phi_{-}|^{2})
+ \lambda^{2}| \Phi_{+}|^{2} | \Phi_{-}|^{2}
 + \frac{g^{2}}{2}
 \left( |\Phi_{+}|^{2}- | \Phi_{-}|^{2} 
+ \xi \right)^{2}
~,\ee
where $Re(S)$ is the gauge singlet inflaton, $\Phi_{\pm}$ are
fields with charges $\pm 1$ with respect to a Fayet-Illiopoulos
$U(1)$ gauge symmetry, $U(1)_{FI}$,
and $\xi > 0$ is the Fayet-Illiopoulos
 term. For $|S| > |S_{c}| = g \sqrt{\xi}/\lambda$,
the minimum of $V(\Phi_{+}, \Phi_{-};|S|)$ is at $\Phi_{\pm} = 0$.
With $\Phi_{\pm} = 0$, the tree-level $S$ potential is flat
 with $V = V_{o} \equiv g^{2}\xi^{2}/2$
 ($\xi^{1/2} \approx 8.5 \times 10^{15} \GeV$ from
COBE normalization \cite{lr}).
One-loop corrections result in a
 potential for $S$ which causes $S$ to slow-roll
 towards $S = 0$ \cite{dti}. Once
 $|S| < |S_{c}|$, the minimum of the potential
for a given value of $|S|$ is at
$\Phi_{+} = 0$ and
\be{e6}  |\Phi_{-}|  =
  \sqrt{ \xi - \frac{\lambda^{2} |S|^{2}}{g^{2}} }    ~.\ee
(In the following we may consider $S$ and $\Phi_{-}$ to be real.) 
Thus the expectation value of $\Phi_{-}$ at the minimum of its potential
is a function of the value of $S$.
The mass squared terms along the $S$, $\Phi_{+}$ and $\Phi_{-}$ directions
as a function of $|S|$ and the $\Phi_{-}$ expectation value \eq{e6} are 
$m_{S}^{2}  = \lambda^{2} |\Phi_{-}|^{2}$, $m_{\Phi_{+}}^{2} =
\lambda^{2} |\Phi_{-}|^{2} + 2 \lambda^{2} |S|^{2}$ and
 $m_{\Phi_{-}^{'}}^{2} = m_{A}^{2} = 2 g^{2} |\Phi_{-}|^{2}$, where
$m_{\Phi_{-}^{'}}$ is the mass 
at $\Phi_{-} \neq 0$ minimum \eq{e6} and $A$ is
 the $U(1)_{FI}$ gauge boson.

\section{Condensate Instability in Hybrid Inflation Models}

              The dependence of the minimum of the $\Phi_{-}$ potential
 on the value of the $S$ field is the reason
 for the instability of the inflaton condensate.
Once $|S| < |S_{c}|$, $S$ and $\Phi_{-}$
oscillate about the minimum of their potentials. Oscillations begin once
$m_{S} > H$. In the case of D-term inflation, this is satisfied once
$|S|^{2}/|S_{c}|^{2} = 1  - 4 \pi g^{2} \xi/3 \lambda^{2} M_{Pl}^{2}
= 1 - 2 \times 10^{-6} g^{2}/\lambda^{2}$ (using $m_{S} = \lambda |\Phi_{-}|$, 
with $|\Phi_{-}|$ as given by \eq{e6}), so  $S$ oscillations typically
begin when $|S|$ is close to  $|S_{c}|$.
 The equation on motion for the inflaton, in terms of the conventionally
 normalized real scalar field $s = \sqrt{2} Re(S)$, is
\be{e7} \ddot{s} + 3 H \dot{s} - \frac{\underline{\nabla}^{2}}{a^{2}}s =
-\lambda^{2} s \left| \Phi_{-} \right|^{2}     ~,\ee
where $H = \dot{a}/a$ is the expansion rate and $a$ is the scale factor.
Suppose we consider the growth of a small spatial perturbation
of $s$. As $s$ decreases below $s_{c}$,
 the mean value of the oscillating $\Phi_{-}$ field
at a point in space will be approximately equal
 to the value at minimum of the $\Phi_{-}$ potential
at that point in space, which depends of
 $s(\vec{x},t)$. So if we average over the
 coherent oscillations of $\Phi_{-}$ about the minimum and replace
 $\Phi_{-}$ by the value $\Phi_{-}(s)$ at the minimum of its potential,
the $s$ equation of motion becomes
\be{e9}  \ddot{s} + 3 H \dot{s} -
\frac{\underline{\nabla}^{2}}{a^{2}}s \approx
-\lambda^{2} \xi s + \frac{\lambda^{4} s^{3}}{2 g^{2}}      ~.\ee
Therefore the $s$ scalar field and perturbations will evolve as if the
 $s$ field had a potential
\be{e10}  V_{eff}(s) \approx \frac{\lambda^{2} \xi s^{2}}{2}  -
\frac{\lambda^{4} s^{4}}{8 g^{2}}        ~.\ee
This is a flatter than $s^{2}$ potential,
 corresponding to an attractive interaction
amongst the $s$ scalars and a negative
 pressure in the condensate \cite{turner,jmcd}.
 Therefore spatial perturbations of the $s$
 condensate will grow, eventually
becoming non-linear and resulting in fragmentation
 into condensate lumps
 \cite{ks,bbb1,bbb2}. The procedure of
 averaging over coherent oscillations of the $\Phi_{-}$ field is
well-defined if $m_{\Phi_{-}^{'}}$ is large
 compared with $m_{S}$, which is true
if $\sqrt{2} g$ is large compared with
 $\lambda$, and we will focus on this case.
 In the case where one cannot first
average over the $\Phi_{-}$ oscillations the
 combined dynamics of the $S$ and
 $\Phi_{-}$ field will be more complicated. A
 particular case of this is F-term hybrid inflation, for
 which there is only a single
 coupling in the scalar potential such that the condition
 $\lambda =\sqrt{2}g$ is
 effectively satisfied \cite{fti,lr}. In this case there
 exists an exact solution of the
 scalar field equations such that the inflaton is
 described by an effective potential of
 the form $a s^{2} - b |s|^{3} + c s^{4}$ ($a,b,c > 0$) \cite{king}.

             Although we have derived 
$V_{eff}(s)$ for the example of D-term inflation, 
we emphasize that a
$-s^{4}$ attractive interaction is a 
generic feature of all hybrid inflation models for
which we can average over the oscillations of the 
field terminating inflation prior to
discussing the dynamics of the inflaton. Therefore our analysis may be readily
applied to other hybrid inflation models. 

\section{Evolution of Perturbations}

   We next consider the growth of spatial perturbations
 and the fragmentation of the inflaton condensate.
 The linear growth of perturbations
has been discussed for a complex scalar field
in the context of Q-ball formation in \cite{ks}, 
using the approach of \cite{lee}.
Here we adapt this approach to the case of
 a real scalar field in the expanding Universe.
 The equation of motion for a real scalar field $\Phi$ is
\be{e12} \ddot{\Phi} + 3 H \dot{\Phi}
 - \frac{\underline{\nabla}^{2}}{a^{2}} \Phi =
- \frac{\partial V(\Phi)}{\partial \Phi}     ~.\ee
We will assume throughout that $V(\Phi)$ is a polynomial with $V(\Phi)
 = V(-\Phi)$. We define $\Phi = (a_{o}/a)^{3/2} \phi$,
 where $a_{o}$ is the scale factor when the
 coherent oscillations begin. The equation of motion then becomes
\be{e13} \ddot{\phi} - \frac{\underline{\nabla}^{2}}{a^{2}} \phi =
- \frac{\partial U(\phi)}{\partial \phi}     ,\ee
where
\be{e13a}   \frac{\partial U(\phi)}{\partial \phi}
  =  \left(\frac{a}{a_{o}}\right)^{3/2}
\frac{\partial V(\Phi)}{\partial \Phi} + \Delta_{H} \phi    ~,\ee
where
\be{e13b} \Delta_{H} = -\frac{3}{2}\left(\dot{H}
+ \frac{3}{2}H^{2}\right)    ~.\ee
With $\phi = R Sin \Omega$, the equation of motion becomes
\be{e14}
(\ddot{R} - R \dot{\Omega}^{2}
 - \frac{\underline{\nabla}^{2}}{a^{2}}R 
+ \frac{R(\partial_{i}\Omega)^{2}}{a^{2}})Sin \Omega +
(-R \ddot{\Omega} -2 \dot{R} \dot{\Omega}
 +\frac{2 \partial_{i}R \partial_{i}
\Omega}{a^{2}} + R
\frac{\underline{\nabla}^{2}}{a^{2}} \Omega) Cos \Omega =
- \frac{\partial U(\phi)}{\partial \phi}    ~,\ee
where $\partial_{i} = \partial/\partial x_{i}$  ($i=1,2,3$).
Multiplying the equation by $Sin \Omega$ and averaging over coherent
oscillations gives
\be{e15}\ddot{R} - R \dot{\Omega}^{2}
 - \frac{\underline{\nabla}^{2}}{a^{2}}R
 + \frac{R(\partial_{i}\Omega)^{2}}{a^{2}} =
-\frac{\partial U_{eff}(R)}{\partial R}  ~,\ee
where
\be{e15a}  \frac{\partial U_{eff}(R)}{\partial R}
 = \frac{2}{R} \left< \phi \frac{\partial U}{\partial \phi} \right>
~,\ee
and where $<f(\Omega)> \equiv \frac{1}{2 \pi} \int_{0}^{2 \pi}
 f(\Omega) \;d \Omega$ denotes averaging over oscillations.
($\Delta_{H}$ is effectively constant on the timescale
 of coherent oscillations. In fact,
 since $\Delta_{H}$ is at most of the order of $H^{2}$, the
 $\Delta_{H}$ term in general
 contributes a negligible correction to the $\phi$ mass squared
 term. Thus we take
$\Delta_{H} = 0$ in the following.)
Multiplying both sides by $Cos \Omega$ and averaging gives
\be{e16} \ddot{\Omega} + 2 \frac{\dot{R}}{R}
 \dot{\Omega} - \frac{2}{R}
\frac{\partial_{i}R \partial_{i}
\Omega}{a^{2}} -
\frac{\underline{\nabla}^{2}}{a^{2}} \Omega = 0    ~.\ee
In this we are assuming that $R$ and 
$\dot{\Omega}$ do not vary much over the period of the oscillations. 
In practice we will be applying this method to 
the case of a $-\Phi^{4}$ interaction
term in the potential.
Therefore this method is accurate if the $-\Phi^{4}$  
term is a small perturbation of the $\Phi^{2}$ term. 
(In the pure $\Phi^{2}$ limit $R$ and $\dot{\Omega}$ are constant.) 

With $R = R + \delta R(\vec{x},t)$ and
$\Omega = \Omega(t) + \delta \Omega(\vec{x},t)$,
the perturbation equations are
\be{e17}  \delta \ddot{R} - \dot{\Omega}^{2}
 \delta R - 2 \dot{\Omega} \delta \dot{\Omega} R -
 \frac{\underline{\nabla}^{2}}{a^{2}}\delta R
 = -\left(\frac{\partial^{2}
U_{eff}}{\partial R^{2}}\right)_{R(t)} \delta R    ~\ee
and
\be{e18} \delta \ddot{\Omega} + 2 \frac{\dot{R}}{R}
\delta \dot{\Omega}
+ 2 \frac{\dot{\Omega}}{R} \delta \dot{R}
-2 \frac{\dot{R} \dot{\Omega}}{R^{2}} \delta R
- \frac{\underline{\nabla}^{2}}{a^{2}}
\delta \Omega = 0   ~.\ee
Assuming the perturbations have the form
$\delta R = \delta R_{o} e^{S(t) -
i \vec{k}.\vec{x}}$, $\delta \Omega
 = \delta \Omega_{o} e^{S(t) -
i \vec{k}.\vec{x}}$ \cite{ks,lee},
the perturbation equations become
\be{e19}  (\dot{\alpha} + \alpha^{2} - \dot{\Omega}^{2}
 + \frac{\vec{k}^{2}}{a^{2}} + U_{eff}^{''})
\delta R = 2 \alpha R \dot{\Omega} \delta \Omega   ~\ee
and
\be{e20}  \left(\dot{\alpha} + \alpha^{2}
+ \frac{\vec{k}^{2}}{a^{2}}
+ 2 \frac{\dot{R}}{R} \alpha \right)\delta \Omega
 = \delta R \left( \frac{2 \dot{R} \dot{\Omega}}{R^{2}}
 - \frac{2 \dot{\Omega}
\alpha}{R} \right)   ~,\ee
where $\alpha = \dot{S}$. Combining
these gives a dispersion relation \cite{lee,ks},
\be{e21}  \left(\dot{\alpha} + \alpha^{2} - \dot{\Omega}^{2}
 + \frac{\vec{k}^{2}}{a^{2}} + U_{eff}^{''}\right)
\left(\dot{\alpha} + \alpha^{2} + \frac{\vec{k}^{2}}{a^{2}} + 2
 \frac{\dot{R}}{R}\alpha \right) =
2 \alpha  R \dot{\Omega} \left(\frac{2 \dot{R}
 \dot{\Omega}}{R^{2}}
 - \frac{2 \dot{\Omega} \alpha}{R}\right)      ~.\ee
In the case where there is no expansion the amplitude of
 oscillation is constant and so we have $\dot{R} = 0$.
A growing perturbation solution is then given by \cite{lee,ks}
\be{e22}  \alpha^{2} =
\frac{\frac{\vec{k}^{2}}{a^{2}} \left( \frac{U_{eff}^{'}}{R}
 - U_{eff}^{''}\right)}{\left(
\frac{3 U_{eff}^{'}}{R} + U_{eff}^{''}\right)}   ~\ee
and $\dot{\alpha} = 0$. 
In deriving this we have used $\dot{\Omega}^{2} = U_{eff}^{'}/R$ 
(from \eq{e15} with constant $R$).
This solution exists if
 $\vec{k}^{2}/a^{2}$ is less than
$\vec{k}^{2}_{max}/a^{2} = \left(
\frac{U_{eff}^{'}}{R} - U_{eff}^{''}\right)$.
(In obtaining \eq{e22} it is assumed that 
$16 (\vec{k}^{2}/a^{2})U_{eff}^{'}/R$ is small compared with
$(3 U_{eff}^{'}/R + U_{eff}^{''})^{2}$, which is satisfied for all 
$\vec{k}$ up to $\vec{k}_{max}$ in the case where the $-\Phi^{4}$ potential 
term is small compared with the $\Phi^{2}$ term.) 

In the case with expansion we generally
 have to solve the equations of motion and
 perturbation equations numerically.
  However, for condensate fragmentation
 we will be mostly interested in the case where
$\vec{k}^{2} =  \vec{k}^{2}_{max}$, corresponding to
 the largest value of $\alpha$ at a given time and so the
 first perturbation mode to go non-linear.
We will also be considering oscillation amplitudes such that the 
$-\Phi^{4}$ potential term is small compared with the $\Phi^{2}$ term.
In this case $R$ may be considered constant throughout. 
Then if $\dot{\alpha}$ is non-zero, the solution $\eq{e21}$ generalizes to
\be{e22a}  \tilde{\alpha}^{2}  =
\frac{\frac{\vec{k}^{2}}{a^{2}} \left( \frac{U_{eff}^{'}}{R} - U_{eff}^{''} 
 \right)}{\left(
\left(4 \gamma - 1\right) 
\frac{U_{eff}^{'}}{R} + U_{eff}^{''}\right)}   ~\ee
where $\tilde{\alpha} = \alpha^{2} + \dot{\alpha}$ and 
$\gamma = \alpha^{2}/\tilde{\alpha}^{2}$. 
Since typically $|\dot{\alpha}/\alpha| 
\approx H$, we see that \eq{e22} will be approximately 
correct for the case $\vec{k}^{2} = \vec{k}_{max}^{2}$ so long as
 $\alpha(\vec{k}_{max}) > \left|\frac{\dot{\alpha}}{\alpha}\right| \approx  H$. 
The solution of the perturbation equations is then 
\be{e23} \delta R \approx \delta R_{o}
 exp\left( \int \alpha dt \right) \; e^{i\vec{k}.\vec{x}}     ~\ee
and
\be{e24} \delta \Omega \approx \delta \Omega_{o}
exp \left( \int \alpha dt\right) \; e^{i\vec{k}.\vec{x}} ~.\ee

      We next apply the above to the case of the generic attractive
 $-\Phi^{4}$ interaction of hybrid inflation models,
\be{e25}  V(\Phi) =  \frac{m^{2} \Phi^{2}}{2}
 - \frac{\eta \Phi^{4}}{4}    ~.\ee
From \eq{e15a},
\be{e26}  \frac{\partial U_{eff}(R)}{\partial R} = \frac{2}{R}
\left< m^{2} \phi^{2}    - \eta \phi^{4}
 \left(\frac{a_{o}}{a}\right)^{3} \right>    =
\frac{2}{R} \left(\frac{m^{2} R^{2}}{2}    - \frac{3 \eta R^{4}}{8}
 \left(\frac{a_{o}}{a}\right)^{3} \right)
  ~,\ee
where we have used $<Sin^{2}\Omega> = 1/2$
 and $<Sin^{4}\Omega> = 3/8$.
Thus
\be{e27} U_{eff}(R) = \frac{m^{2} R^{2}}{2}
 - \frac{3 \eta R^{4}}{16}
 \left(\frac{a_{o}}{a}\right)^{3}     ~.\ee
Thus from \eq{e22} we find
\be{e28} \alpha =  \left(\frac{a_{o}}{a}\right)^{3/2}
\left(\frac{3 \eta R^{2}}{8 m^{2}}
\right)^{1/2}
 \left(1 - \frac{9 \eta R^{2}}{8 m^{2}}
\left(\frac{a_{o}}{a}\right)^{3} \right)^{-1/2} \frac{|\vec{k}|}{a}   ~.\ee
\eq{e28} is strictly valid only 
if the $-\Phi^{4}$ term is a small perturbation of the $\Phi^{2}$
term. In this case $R$ will be essentially 
constant (equal to its initial value $R_{o}$) and 
$\left(1 - \frac{9 \eta R^{2}}{8 m^{2}}
\left(\frac{a_{o}}{a}\right)^{3} \right)$
will be approximately equal to 1, conditions which we will 
assume to be satisfied 
in the following. 

         The growth of the perturbations is then given by,
\be{e29} \frac{\delta R}{R} \approx
\frac{\delta R_{o}}{R_{o}} \;
 exp\left[ \left(\frac{3 \eta R_{o}^{2}}{8 m^{2}}
\right)^{1/2} \frac{|\vec{k}|}{aH_{o}}
\left(\frac{2}{5-2n}\right)
\left(\frac{a_{o}}{a}\right)^{3/2-n} \left(
\left(\frac{a}{a_{o}}\right)^{5/2-n} - 1\right)  \right]  ~.\ee
In this we have used $H = H_{o}(a_{o}/a)^{n}$, where
 $n$ will be between 0
and 3/2 as the $\Phi$ oscillations develop
 from the end of inflation to
an approximately $\Phi^{2}$ potential.
The condition for fragmentation to occur
 is that $\delta R/R \gae 1$.
The largest growth at a given time corresponds to the
 mode $\vec{k}_{max}$,
where
\be{e29a} \frac{\vec{k}_{max}^{2}}{a^{2}} =
\left( \frac{U_{eff}^{'}}{R} - U_{eff}^{''}\right)
= \frac{3}{2} \eta R^{2} \left(\frac{a_{o}}{a}\right)^{3}    ~.\ee
This determines the radius of the condensate lumps
 when the condensate fragments,
$r_{l}$,
\be{f39} r_{l} \approx \frac{\pi a}{|\vec{k}_{max}|}  =
\left(\frac{2}{3}\right)^{1/2}
\left(\frac{a}{a_{o}}\right)^{3/2}
 \frac{\pi}{\left(\eta R^{2}\right)^{1/2}}
~.\ee
This is really the initial radius of the lumps
 immediately after fragmentation, and the lump
will subsequently relax to its stable configuration,
in which the attractive potential term is
balanced by the gradient term in the equation of
 motion. However, the radius of the stable
 configuration is similar to that of the initial
 lump, since for a stable configuration of the form
 $\phi(r,t) \approx \phi(r) Sin(mt)$, the equation for $\phi(r)$ is
$\partial^{2}\phi(r)/\partial r^{2} + (2/r) \partial
 \phi(r) \partial r \approx Sin^{-1}(mt)\partial
 \delta V /\partial \phi$, where $V = m^{2} \phi^{2}/2
 + \delta V$. For a stable lump of radius $r_{s}$ and
 field amplitude $\phi$ we therefore expect
 that $r_{s}^{2} \approx \phi |\partial \delta
 V /\partial \phi|^{-1}$ (the left hand side
 of the $\phi(r)$ stable lump equation being $\sim \phi/r^{2}$). For
 $\delta V = -\eta \phi^{4}/4$ and $\phi \approx R$
 this implies that  $r_{s} \approx (\eta R^{2})^{-1/2}$.

    To give a condition for inflaton
 condensate fragmentation,
we use $\vec{k} = \vec{k}_{max}$.
The condition for fragmentation to occur is then
\be{e30}
\frac{1}{2 m} \left(\frac{3 \eta R_{o}^{2}}{2}\right)
\frac{1}{H_{o}} \left(\frac{2}{5 - 2n}\right)
\left( \left(\frac{a_{o}}{a}\right)^{1/2}
- \left(\frac{a_{o}}{a}\right)^{3-n} \right)
 \gae \beta
 \equiv log\left(\frac{R_{o}}{\delta R_{o}}\right)   ~.\ee

As $a$ increases, the left hand side of \eq{e30} is
maximized for $\frac{a_{o}}{a}
 = (\frac{1}{2\left(3-n\right)})^{\frac{2}{5-2n}}$
 and then decreases.
Condensate fragmentation in hybrid inflation models
must therefore occur soon after coherent oscillations
begin if it is to occur at all. 

    Assuming that the $-\Phi^{4}$ contribution to the potential is small, 
the coherent oscillations are approximately $\Phi^{2}$ and so 
$n=3/2$, such that maximum growth occurs at $\frac{a_{o}}{a} = \frac{1}{3}$. 
The condition for condensate fragmentation is then
\be{e31} \frac{1}{2 \sqrt{3}}
 \frac{\eta R_{o}^{2}}{m H_{o}}
\gae \beta   ~,\ee
where $H_{o}$ is calculated using for the energy density 
$\rho = m^{2} R_{o}^{2}/2$. 

\section{Application to D-term Inflation}

               For the case of D-term inflation,
$m^{2} = \lambda^{2} \xi$ and $\eta = \lambda^{4}/2 g^{2}$. In order that 
the $-\Phi^{4}$ contribution to the potential is small, 
we choose $R_{o} = s_{c}/2$, such that $R_{o}^{2} =  g^{2} \xi/2\lambda^{2}$.
(The perturbations will start to grow as soon as coherent 
oscillations begin with amplitude  $s \approx
 s_{c}$. However, as we cannot apply our method to calculate 
the growth for oscillation amplitudes between $s_{c}$ and $s_{c}/2$, 
we are in fact {\it underestimating} the total growth.) 
Thus from \eq{e31} the condition
 for condensate fragmentation
becomes
\be{e32} \frac{\lambda}{g} \frac{1}{\beta} \gae
\frac{8 \sqrt{2 \pi}
\xi^{1/2}}{M_{Pl}} \approx 0.014
  ~.\ee
In order to complete the fragmentation condition we need the value of
$\beta$.  The seed perturbations of the inflaton field are expected
to come from quantum de Sitter fluctuations of the
 scalar field during inflation. 
Modes with wavenumber $k$ large compared w
ith $H$ will be excited by the increase 
of the inflaton mass at the end of inflation from approximately zero to $m_{S} 
\approx \lambda \xi^{1/2}$. The largest wavenumber 
excited typically corresponds to 
$k_{m} \approx m_{S}$, with amplitude 
$\delta s \approx m_{S}/(2 \pi)$ \cite{lyth}.
Since $|\vec{k}_{max}|/a \approx \sqrt{3/2} \lambda \xi^{1/2} \approx k_{m}$ 
at $a \approx a_{o}$, 
seed perturbations of wavenumber $\vec{k}_{max}/a$ will 
exist after inflation ends, with
\be{e34} \beta \approx log\left(\frac{2 \pi s_{c}}{m_{S}}\right)
= log\left(\frac{2 \sqrt{2} \pi g}{\lambda^{2}}\right) ~,\ee
where we have used $\delta R_{o}/R_{o} \approx \delta s/s_{c}$. 
So with $\lambda, g \gae 0.1$, $\beta \approx 5 - 10$ is a typical value. 
Therefore, using $\beta = 10$, we find that the 
condition for fragmentation to occur is 
\be{e34x}  \lambda \gae 0.2 g    ~.\ee
Since this neglects the growth of perturbations 
for oscillation amplitudes between $s_{c}$
 and $s_{c}/2$, the true lower bound on $\lambda$ 
for fragmentation is likely to be smaller. 
The radius of the condensate lump relative to
 the horizon radius when the condensate fragments is then
\be{x1} \frac{r_{l}}{H^{-1}} \approx
\left(\frac{8 \pi^{3}}{9}\right)^{1/2}
\frac{g}{\lambda}
\frac{\xi^{1/2}}{M_{Pl}}
= 3.7 \times 10^{-3}\; \frac{g}{\lambda}
~.\ee
The condition for the approximations
leading to \eq{e23} to be consistent,
$\alpha(\vec{k}_{max}) > \left|\frac{\dot{\alpha}}{\alpha}\right|  
= 5H/2$ at $\frac{a_{o}}{a} = 
\frac{1}{3}$ (where we assume $H \propto a^{-3/2}$ 
for $s < s_{c}/2$ and we have used $\alpha \propto a^{-5/2}$ from \eq{e28}), 
is satisfied if
\be{e34a} \frac{\lambda}{g} > 
\frac{40 \sqrt{2 \pi}\xi^{1/2}}{M_{Pl}} \approx 0.07 ~.\ee
So if the fragmentation condition \eq{e34x} is satisfied then the approximations
 are consistent.

             Therefore in
 D-term inflation models condensate fragmentation is likely to occur if
$\lambda \gae 0.2g$. Since the true lower bound on $\lambda$ for fragmentation 
is likely to be smaller, it is probable
 that condensate fragmentation will occur
if $\lambda$ and $g$ both take values in the natural range 0.1-1.

\section{Enhancement of Inflaton Annihilations By Condensate Fragmentation}

                   We next consider the possible
 consequences of inflaton condenate
fragmentation. One potentially important consequence
 is that inflaton annihilations are
effectively enhanced compared with the case of a homogeneous
 condensate and may
dominate over decays as the primary mode of reheating.

          We first show that once the condensate fragments
 the particles within the
condensate lump
decouple from the expansion of the Universe, such
 that the number density
 and field amplitude inside the condensate lumps remains constant.
This will be true if the force on the particles due to scalar
 interactions is greater than
the gravitational force responsible for slowing the
 expansion of the Universe. Suppose the
energy density is dominated by a pressureless
 homogeneous energy density
 $\overline{\rho}$.
(For the case of inflaton condensate fragmentation
 this is not strictly true, since $\delta
 \rho/\overline{\rho} \approx 1$ when the condensate
 fragments. However, it is useful to make this
 assumption in order to obtain an expression that can
 be applied to scalar field models in general.) Suppose then we
consider a spherical lump of radius $r$. The gravitational
 acceleration acting on a particle at the surface of the lump is then
\be{f35} \ddot{r} = - \frac{4 \pi \overline{\rho}}{3 M_{Pl}^{2}}r    ~.\ee
If the force due to the attractive scalar interaction produces a smaller
acceleration than this, the particles will follow the
 expansion of the Universe, otherwise they will
 decouple from expansion. To estimate the force due
 to the scalar interaction consider a
sphere of radius $r$ and with a fixed number of
 scalar particles $N$,
\be{f35a} N = \frac{4 \pi r^{3}}{3}
\left(\frac{m \phi^{2}}{2}\right)     ~,\ee
where the number density of scalars is $n = m \phi^{2}/2$
and where for simplicity we have considered a constant
 amplitude $\phi$
for the coherently oscillating field inside
the sphere. This gives $\phi$ as a function of $r$.
With $V(\phi) = \frac{m^{2} \phi^{2}}{2} + \delta V$,
the total potential energy of the sphere is
\be{f36} E_{pot} = \frac{4 \pi r^{3}}{3}
 \left( \frac{m^{2}\phi^{2}}{2}
+ \delta V \right)  = mN  + \frac{4 \pi r^{3}}{3} \delta V~.\ee
Thus the force due to the scalars is
\be{f37} F_{s} = -\frac{dE_{pot}}{dr} = -\left(4 \pi r^{2} \delta V -
\frac{6N}{mr} \frac{\partial \delta V}{\partial \phi^{2}}\right)    ~.\ee
For the case $\delta V  = - \eta \phi^{4}/4$, the force is
$F_{s} = - \pi \eta \phi^{4} r^{2}   $.
The condition that the acceleration due to
 scalar attraction is larger than the
gravitational acceleration is then
\be{f38} \phi^{4} > \frac{m H^{2}}{2 \pi \eta r}    ~.\ee
Since condensate fragmentation occurs soon after
coherent oscillations begin
(at $\partial V/\partial \phi \approx 0$), we
 have $\phi^{2} \approx m^{2}/\eta$. With
$r \approx r_{l}$  (with $a \approx a_{o}$ in \eq{f39})
 \eq{f38} becomes
\be{f38a}  \frac{m^{3}}{\eta^{3/2}R} >
\left(\frac{3}{2}\right)^{1/2}
\frac{H^{2}}{2 \pi^{2}}     ~.\ee
Note that if this condition is satisfied then from \eq{f35} we have 
$|\ddot{r}/r| \gae H^{2}$, which can be rewritten as $|\ddot{r}|H^{-1} \gae 
Hr \equiv \dot{r}$. Therefore the attractive 
force between the scalars will bring the 
expansion of the lump to a halt within an 
expansion time, $\delta t \approx H^{-1}$. 
For the case of D-term inflation, \eq{f38a} is satisfied if
\be{f40} \lambda < 
\left(\frac{\sqrt{6} \pi M_{Pl}^{2}}{\xi}\right)^{1/2}
\approx  4 \times 10^{3} ~,\ee
which is generally strongly satisfied. Thus the scalars
 in the condensate
lumps decouple from the gravitational expansion.

       To see how condensate fragmentation effectively
enhances annihilations, consider an interaction
$\lambda^{2} |S|^{2} |Q_{i}|^{2}$ between inflatons
 and light scalars $Q_{i}$ (in D-term inflation $Q_{i}$
 will correspond to MSSM scalars \cite{kmr}).
We will calculate the annihilation rate assuming no Bose enhancement
(i.e. no parametric resonant decay of the condensate),
in which case the annihilation rate simply corresponds to the perturbative
annihilation rate of the scalars in the condensate.
The average of the annihilation cross-section times relative particle
 velocity in the $v \rightarrow 0$ limit is \cite{jgs}
\be{e39} <\sigma v>_{ann} = \frac{\lambda^{4}}{64 \pi m_{S}^{2}}    ~.\ee
With the $s$ number density in the condensate given by
$n = m_{S} s^{2}/2$, the annihilation rate of scalars in the condensate is then
\be{e40}   \Gamma_{ann} = n <\sigma v>_{ann}
 = \frac{\lambda^{4}s^{2}}{128 \pi m_{s}}   ~.\ee
The condition for scalars to annihilate is then
 $\Gamma_{ann} > H \equiv H_{o}(a_{o}/{a})^{3/2}$.
 We can now see why annihilation is effectively
 enhanced when the condensate fragments.
 In the case of a homogeneous condensate, the
 scalars are freely expanding and
 $s^{2} \propto a^{-3}$, so the annihilation
 rate drops more rapidly than the expansion
 rate as the scale factor $a$ increases. Thus unless
 annihilations are effective immediately
after the end of inflation, they will never
 be significant and inflaton decays will be the main
mode of reheating. However, if the inflaton
 condensate fragments soon 
after oscillations begin then the value of the
 amplitude $s$ inside the lumps is {\it constant} and so
 $\Gamma_{ann}/H \propto a^{3/2}$ {\it increases} as $a$ increases.
 Therefore as the Universe expands annihilations
 will eventually occur and may be the dominant
 mode of reheating.  The reheating temperature
 due to annihilations is then
\be{e44} T_{R} =
  \left(\frac{45}{4 \pi^{3} g(T_{d})}\right)^{1/4}
\left(\Gamma_{d} M_{Pl}\right)^{1/2}
= \left(\frac{1}{128 \pi}\right)^{1/2}
\left(\frac{M_{Pl}}{m_{S}}\right)^{1/2}
\left(\frac{45}{4 \pi^{3} g(T_{d})}\right)^{1/4}
\lambda^{2} s_{c}     ~,\ee
where we have used $s \approx s_{c}$ inside the lumps and
where $g(T_{d})$ is the number of degrees of
 freedom in thermal equilibrium \cite{eu}.

     We next apply this to the specific case
 of D-term inflation. If the MSSM fields also
carry $U(1)_{FI}$ charges, there is an
 interaction with the scalars of the MSSM
of the form $\lambda^{2} |Q_{i}|^{2} |S|^{2}$
coming  from
 integrating out the $U(1)_{FI}$ gauge fields
 \cite{kmr}, where $\lambda$ is the
 superpotential coupling from \eq{e3}.
The reheating temperature is then
\be{e45} T_{R} = \left(\frac{1}{64 \pi}\right)^{1/2}
\left(\frac{45}{4 \pi^{3} g\left(T_{d}\right)}\right)^{1/4}
g \lambda^{1/2} \xi^{1/4}M_{Pl}^{1/2}
\approx 5.5 \times 10^{15}
\left(\frac{100}{g(T_{R})}\right)^{1/4} g \lambda^{1/2}
\GeV   ~.\ee
The upper bound from requirng that gravitinos
 are not generated excessively
by thermal scattering is $T_{R} \lae 10^{8-9} \GeV$ \cite{grav}.
Thus we require
\be{e45a} g \lambda^{1/2}  \lae 1.8 \times 10^{-7}
\left(\frac{g(T_{R})}{100}\right)^{1/4}
\left(\frac{T_{R}}{10^{9} \GeV}\right)
~.\ee
The smallest value of $\lambda$ for which fragmentation
 occurs is $\lambda
\approx 0.2g$. In this case if condensate fragmentation occurs
then from \eq{e45a} the $U(1)_{FI}$ gauge coupling must satisfy $
g \lae 7 \times 10^{-5}$. (This upper bound will be
 even stronger in the presence
of parametric resonance.) This will not be satisfied
 if, for example, the $U(1)_{FI}$
gauge coupling has the typical magnitude $g \approx 1$
 of the MSSM gauge couplings.
Thus condensate fragmentation is typically not
 compatible with D-term inflation
if the inflatons can annihilate to MSSM fields.
In order to have D-term inflation consistent
 with the absence of thermal gravitinos
we must either eliminate inflaton
 annihilations, which requires that
the MSSM fields do not carry $U(1)_{FI}$
 charges, or
eliminate inflaton condensate
 fragmentation, which requires that
$\lambda < 0.2g$.

\section{Inflaton Condensate Fragmentation and
 Tachyonic Preheating}

               Throughout the preceeding discussion 
we have assumed that a homogeneous 
inflaton condensate forms at the end of inflation. 
 However, it has been shown that spatial perturbations of 
the inflaton field may 
grow and become non-linear much more rapidly (before 
homogeneous oscillations are established)
 in a process known as tachyonic preheating 
\cite{tp,tp2}. In this case the final 
state is composed of colliding scalar field waves \cite{tp,tp2}.
Recently it has also been shown \cite{cpr} that condensate lumps occur in 
tachyonic preheating (called "oscillating hot spots" in \cite{cpr}).
Tachyonic preheating typically occurs in less than 
the time for a single coherent oscillation
\cite{tp}, making it impossible to average over a 
coherently oscillating inflaton field. 
Inflaton condensate fragmentation and tachyonic preheating may be regarded as 
different manifestations of a general phenomena, namely the instability of the 
inflaton field with respect to spatial 
perturbations in hybrid inflation models. 
 
    The question of whether inflaton condensate fragmentation or tachyonic
preheating occurs at the end of hybrid inflation will  
depend upon the state of the field at the time when the inflaton reaches 
$s_{c}$, in particular the rate at which the homogeneous field $s(t)$ 
is rolling  (where $s = s(t) + \delta s(\vec{x},t)$) 
relative to the rate of tachyonic growth of the spatial 
perturbations $\delta s(\vec{x},t)$ 
at $s < s_{c}$. If the 
homogeneous field 
can catch up with the spatial perturbations (which 
cross $s_{c}$ before the homogeneous field) 
before there is significant growth of the perturbations due 
to the tachyonic mass term at $s < s_{c}$, 
then there will be no tachyonic preheating. This  
depends crucially on the rate of rolling of the homogeneous field at $s_{c}$, 
which in turn requires that the full inflaton potential with radiative 
corrections be considered. For the case of D-term inflation a
full analysis has yet to be done \cite{mb}. 

\section{Conclusions}

     We have shown that it is possible for the
 inflaton condensate in hybrid inflation models
to fragment to condensate lumps. The inflaton
 condensate is in general unstable, but the
instability reduces as the Universe expands, requiring
 that fragmentation occurs shortly after the end of inflation.
 The state of the Universe after
 fragmentation, with the energy density concentrated
 inside inflaton condensate lumps,
 is quite different from the conventional
 post-inflation scenario of a homogeneous inflaton
 condensate.
One consequence of inflaton condensate
 fragmentation is that
inflaton annihilations will be effectively
 enhanced relative to the case of a
homogeneous inflaton condensate. In the case
 of D-term inflation models, which we have used as a specific
 example of hybrid inflation models in our discussion, if 
condensate formation and fragmentation
 occurs then the enhancement of inflaton
 annihilations implies that the reheating
 temperature is typically large compared with the
 thermal gravitino upper bound. Thus
 either
inflaton annihilations must be suppressed, which requires
 that the MSSM fields do not
 carry $U(1)_{FI}$ gauge charges, or fragmentation
 must not occur, which requires that
 $\lambda < 0.2 g$.

         If inflaton annihilations do not occur, or if
 we consider a non-SUSY model for
 which there is no thermal gravitino bound
 on the reheating temperature, then inflaton
 condensate fragmentation can safely
 occur. This may have interesting consequences
for the inflaton dominated period following
 inflation. For example, the fact that the
energy density of the Universe is now
 concentrated in condensate lumps could greatly alter the dynamics
 of SUSY flat direction scalar fields, which
 acquire a mass from the SUSY
 breaking inflaton energy density in the case
 of a homogeneous inflaton condensate \cite{drt}.
 Another potentially important effect would be
 for parametric resonant decay of the inflaton
 and preheating. In the case of a homogeneous
 condensate, parametric resonance turns off as
 the Universe expands and the inflaton oscillation
 amplitude decreases \cite{add,zlatev}.
However, if the condensate fragments then, just
 as in our discussion of perturbative annihilations,
the oscillation amplitude will be frozen inside the
 lumps and so parametric resonant decay should
 continue without stopping, resulting in more efficient preheating.

   Growth of spatial perturbations of the inflaton field in hybrid inflation 
models has also been demonstrated in the context of 
tachyonic preheating. In general,
the mode by which hybrid inflation ends is likely to be sensitive to both
the model paramaters and the initial conditions at the end of inflation. 
A detailed numerical investigation will therefore be necessary in order to 
clarify how hybrid inflation ends in a given model \cite{mb}. 
However, it should be emphasized that condensate lumps are a feature of 
both inflaton condensate fragmentation 
and tachyonic preheating \cite{cpr}. Therefore our discussion of the 
cosmology of condensate lumps 
should apply to the tachyonic preheating case also. 

       It is important to emphasize that inflaton
 condensate fragmentation is a
natural possibility in all hybrid inflation models
 and that it may have consequences
for cosmology beyond what has been discussed
 here. A detailed understanding of
inflaton condensate fragmentation will
 therefore be necessary in order to
fully understand the cosmology of hybrid inflation models.

\end{document}